# Deep Residual Network based food recognition for enhanced Augmented Reality application

Siddarth Sairaj[1], Sainath Ganesh[2], Vignesh S[3]

[1,2,3] Dept. of Computer Science and Engineering

Vellore institute of Technology, Chennai , India

[1]siddarth.sairaj2017@vitstudent.ac.in, [2]sainath.g2017@vitstudent.ac.in, [3]s.vignesh2017a@vitstudent.ac.in

*Abstract*

*Deep neural network based learning approaches are widely utilized for image classification or object detection based problems with remarkable outcomes. Realtime Object state estimation of objects can be used to track and estimate the features that the object of the current frame possesses without causing any significant delay and misclassification. A system that can detect the features of such objects in the present state from camera images can be used to enhance the application of Augmented Reality for improving user experience and delivering information in a much perceptual way. The focus behind this paper is to determine the most suitable model to create a low-latency assistance AR to aid users by providing them nutritional information of the food that they consume in order to promote healthier life choices. Hence the dataset has been collected and acquired in such a manner, and we conduct various tests in order to identify the most suitable DNN in terms of performance and complexity and establish a system that renders such information realtime to the user.*

## 1. Introduction

Food has become a commodity in today's world. Food is not just a requirement for survival but is also something that pleases us and gives us pleasure. Consuming the required amount and type of food becomes important as food and nutrition requirements differ from one individual to another. There is a need for technology for the people who would like to know about the ingredients their food is made of, as their food choices for each individual might differ based on their personal nutritional requirements and dietary conditions. This paper is aimed at providing a solution for that and in the process, satisfying all food enthusiasts and users who desire to lead a healthier lifestyle by consuming in a smart and healthier way. Monitoring of one's personal health through a mobile device and the idea of restaurants functioning without waiters are real-world examples of where automatic recognition of food items could come into play. A mobile-based AR application of food recognition can help users maintain a check over their daily calorie consumption. A mechanism to estimate nutritional information of foods instantly is not far from reality. In reality, however, automatic food classification systems are yet to be fully adopted and integrated into applications. We have made an attempt at driving the process forward by making use of a dataset that largely consists of Indian food items and training the images with multiple neural network architectures to find out the best one suited for our dataset and application.

## 2. Dataset Preparation

A Food database of strictly Indian foods was created for the purpose of this application, and this is labeled the Food20 dataset. The foods chosen for this were mostly based on the nutrition guidelines recommended and widely consumed food items. This data was mainly scraped from image search engines such as google and yahoo and also from popular photo-sharing application Instagram. Queries were made to these applications so as to filter out appropriate food items and their images. The collected images were then manually examined and filtered out to reduce the number of noisy images present in them. The dataset consists of 2,000 Indian food images and has approximately a minimum of around 100 sample images for each food item class. The acquired images have resolutions in the range of a minimum of 200x150 to 5760x3840 pixels per image. The data has been collected from real-world images and is subject to distortions and improper illuminations of certain regions. A subset of the images acquired also contains multiple food items. Choosing a label for that image becomes a decision based on the food item that occupies the maximum area among all the items present in that image. Images belonging to a certain class or label also have high variability in appearance. The DCNN based ResNet model has then been retrained on this dataset for the purpose of food classification. 70% of the images contributed to the training set, and the remaining 30% was used for the purpose of testing. A cross-validation based approach was followed for the proper use of available data and to improve the accuracy of the model.

## 3. Literature Survey

Ashutosh Singla et al., 2016 have provided a system using a pre-trained GoogleNet classifier for identification and segregating food images from non-food images and also to distinguish the food item into its category or class. They have trained their model on two publicly available datasets and also data gathered from social media. They have further fine-tuned and made improvements to the existing model for achieving higher accuracy [1].

Kuang-Huei Lee et al., 2018 have provided selected examples of CleanNet results on the Food-101 dataset. In this paper, the difficulties in scalability and classification of data with high noise have been illustrated. A transfer-learning based approach has been applied for the training of the CNN model [2].

Amaia Salvador et al., 2019 have proposed an image-to-recipe generation system, which takes a food image as input to produce a recipe consisting of a title, ingredients, and sequence of cooking instructions. The ingredients were predicted using the food images uploaded by the user, and the dependency on modeling was highlighted [3].

Christian Szegedy et al., 2017 have presented a new Inception-v4 model, which is a variation of the Inception model without residual connections that produces similar recognition performance as Inception-ResNet-v2. The introduction of residual connections in the architecture was found to drastically improve the performance of the inception model [4].

Marc Bolaños and Petia Radeva, 2016 have proposed a system that performs both localization and recognition of the food item. The system has been proven applicable for both conventional and noisy images with considerable accuracy achieved for both [5].

Zifeng Wu et al., 2019 have analyzed the architecture ResNet model in terms of the ensemble classifiers and the residual unit depths present in it. A system which is spatially and performance-wise more efficient has been proposed by them for large networks. They have used a bunch of shallow networks over deeper networks for this and have shown that the former achieves better performance over the latter [6].

Kiyoharu Aizawa et al., 2019 have proposed a personalized system for classification of food image recognition in a real-world scenario. In order to achieve better performance, a weight optimization-based algorithm has been employed [7].

Takuya Akiba et al., 2017 in their paper, demonstrate that training the ImageNet dataset on a Resnet-50 DCNN model can take as little as fifteen minutes. They achieved with the help of a 1024 Tesla p100 GPU's, training the model for a total of ninety epochs with an image batch size of thirty-two thousand images. They have used several techniques such as a slow start learning rate scheduler along with batch normalization and RMSprop warmup for achieving such quick training [8].

Krzysztof Wołk et al., 2019 have proposed a system that uses the Resnet-34 neural network model for the classification of images. They have made of the use of Google's cloud collab platform along with fast.ai making it really easy to train a model. They have trained their model on a dataset containing images of 7000 od polish cars and have managed to get an accuracy of 99.18% [9].

Francesco Ragusa et al., 2016 have found out that a combination fine-tuned Alex net model and a binary SVM classifier using the Caffe framework gives the highest accuracy of classifying food and non-food items. They have used the UNICT-FD889, Flickr-Food, and Flickr-NonFood dataset for training purposes [10].

Gianluigi Ciocca et al., 2016 have built a dataset of 1,027 canteen trays with a total of 3,616 food instances belonging to over 73 food classes. They have manually divided the food in each tray using polygonal boundaries. An accuracy of about 79% was achieved by them for food and tray recognition in images after training their dataset using CNN based model [11].

Giovanni Maria Farinella et al., 2016 have built a dataset of 4754 food images of over 1200 unique dishes collected during meals. Each meal plate was collected multiple times. The images of the dataset have been labeled into eight different categories. They have also proposed Anti-Textons way of representation of images in which the spatial information present in images is encoded between many Textons [12].

Tushar Gupta, Mudita Sisodia, has built a MAR prototype using augmented reality based text generation to improve the reading time taken for people suffering from Dyslexia with improved accessibility [13].

Jeongeun Lee designed an augmented reality application for gamification of local food to increase consumption by tourists, and they have documented the test results, which show it can enhance user experience[14].

A.-M. Boutsi1, C. Ioannidis1 , S. Soile1 proposed an AR system that consisted of a system for perception of photogrammetric 3D overlays, monitoring and administration, and JavaScript modules for cross-platform support. The acknowledged model demonstrates that a minimal effort AR work process with open-source segments can serve the most of utilization cases which includes vision-based and hybrid tracking method[15]

## 4. Hardware Specifications

**Image classifier model training(Google Collaboratory)**
1.GPU: 1xTesla K80, compute 3.7, having 2496 CUDA cores , 12GB GDDR5 VRAM
2.CPU: 1xsingle core hyperthreaded Xeon Processors,1 core, 2 threads.

**Hosting Platform**
Google Cloud Platform cloud computing infrastructure.

**Augmented Reality Support**
1.Apple: IOS 10.0 or newer
2.Android 4.1 (API 16) or newer.

## 5. Performance Analysis and Comparison

We have experimented with various state of the art architectures on our food dataset, and have visualized the performances of each algorithm, with classification rate, memory footprint, latency and overall performance as a measure we can find which model learns at a faster rate and outperforms the rest while offering an optimal complexity to perform. The rate at which loss gets altered over batches processed, as seen in Fig 1. The same tests have been taken across multiple models and the figures in this section are indexed in the following order, row1 from left to right:vgg19,vgg16,squeezenet1_0,resnet152 ; row 2 from left to right resnet50,densenet121,alexnet and the results derived from the insights of these tests and other benchmarks are summarized in the Results section to find a suitable model.

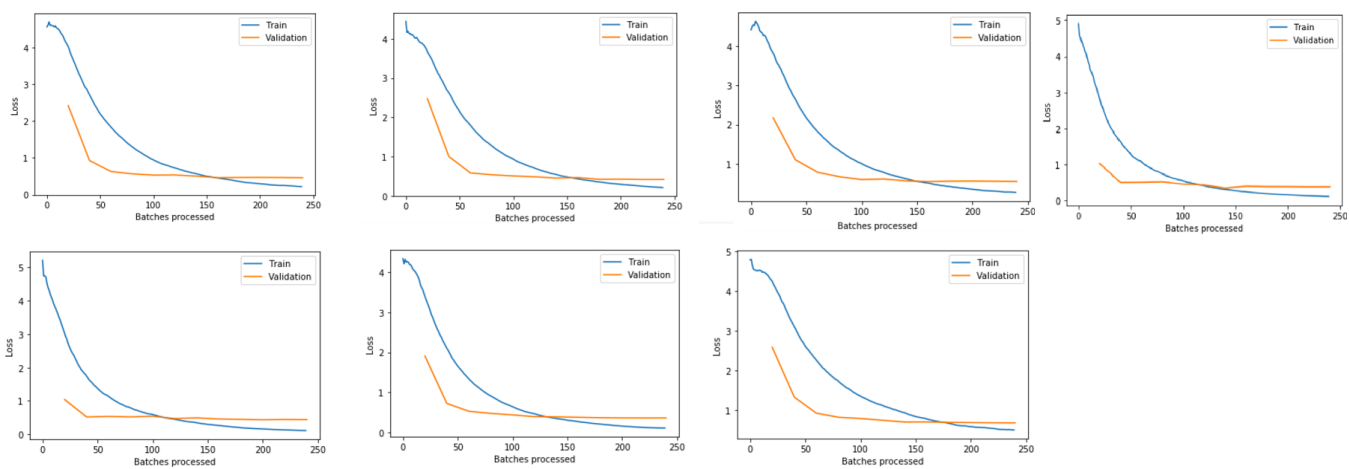

**Fig 1:** Batch loss plot

The learning rate is one of the most important hyperparameters that needs to be configured to modify the model behavior to improve performance and reducing loss, and this can be visualized in Fig 2.

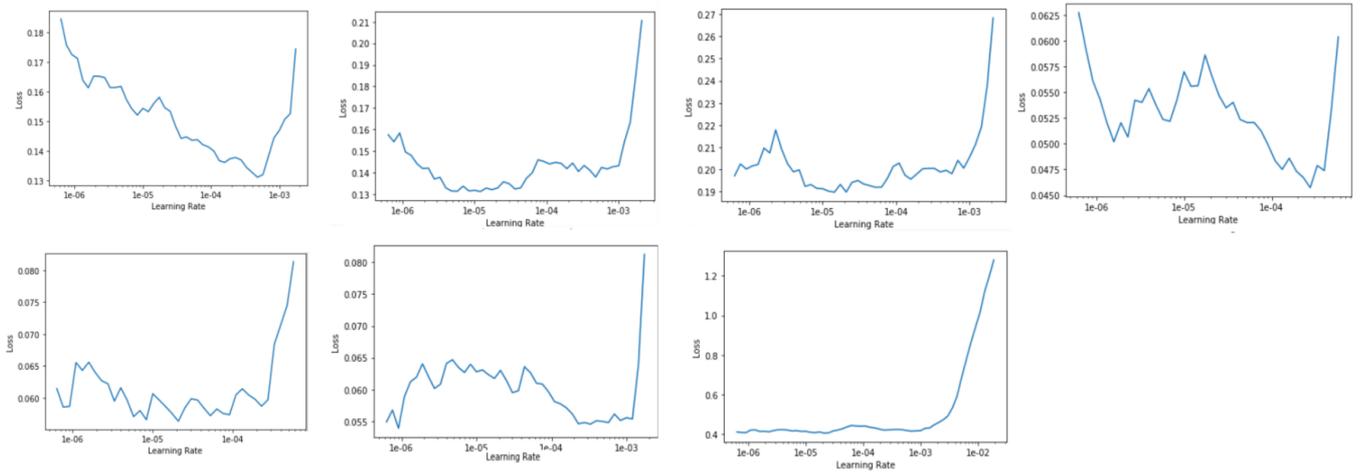

**Fig 2**: Loss over Learning Rate plot

The model's performance measurement can be summarized over actual and predicted classes and gives a clear insight into the types of an error made by the classifier using the confusion matrix, as seen in Fig 3.

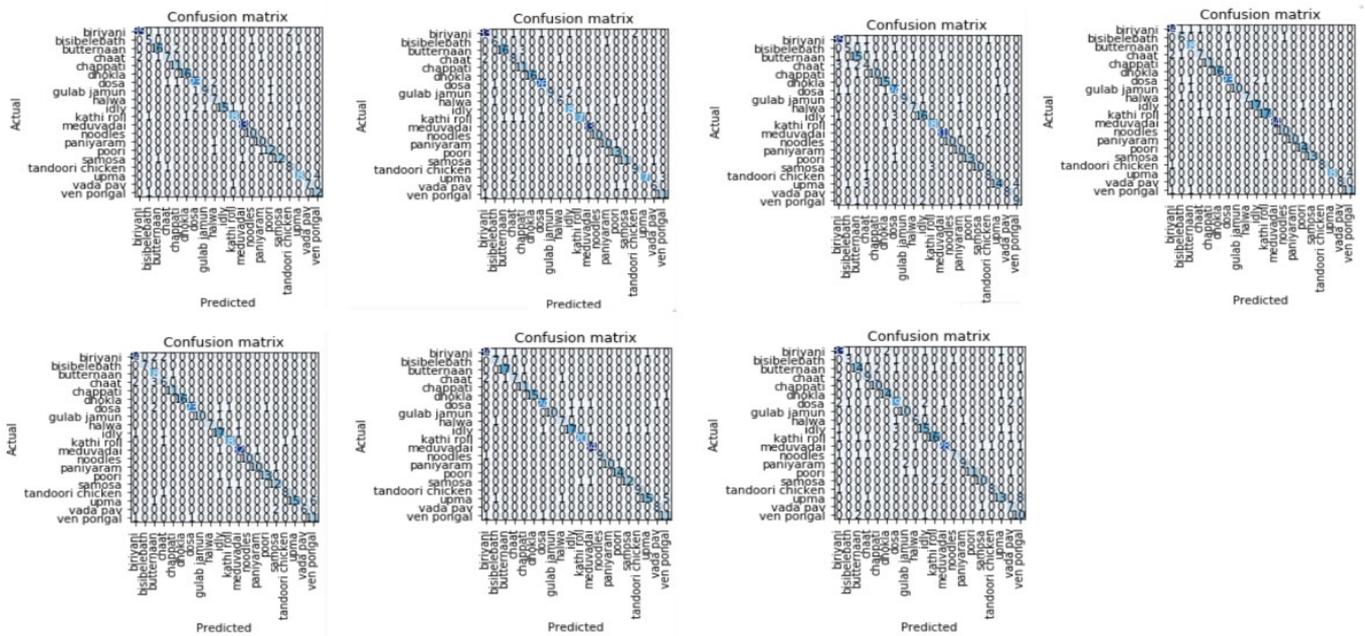

**Fig 3:** Confusion Matrix plot

## 6. Proposed System

Our system combines the use of residual deep neural-network to create a food recognition API with an AR and Rendering engine that encapsulates the use of Image Processing and Asynchronous methodology to display nutritional info estimated. ResNet residual neural network architecture was found to be efficient for testing out our food20-image dataset when compared with widely accepted models. ResNets architecture allows the flexibility to be modified to have variable sizes based on the number of layers sufficient by the system.

**Food Recognition Engine**

ResNet residual neural network architecture was found to be the most efficient for testing out our food dataset. Based on the number of layers in the model, ResNets can be modified to have variable sizes ResNets use ReLU (Rectified Linear Activation Unit) that allows the neural network to skip over other layers of the model that may contain non-linearities. ReLU is basically a non-linear activation function that acts as a linear function and allows the learning of complex data relationships. There are several model optimization techniques for Resnet, which can improve performance to achieve higher accuracy. We have used one of the variants of ResNet, namely ResNet50.

**AR and Rendering Engine**

ReactNative Framework is used for real-time overlaying of food information generated by our food recognition engine on to the real world. A two way API based connectivity has been created between two systems to transfer food and nutritional data, repeated fetch and request method has been implemented with a default time buffer that captures an instance of camera feed frame to receive information from the recognition engine. The camera component is defined that captures data from the real world. React native-modal extension offers a flexible, customizable animation rendering option. A prop is defined and waits for the server to receive a response. Once the information is received, the render function renders the fetched data to the user by augmenting it to the real world. If the camera frame is altered to a different food object or environment, the prop is closed, the rendered data will be removed and will be set back to defaults.

The architecture of the working model is as seen in Fig 4. Once the user uses the application, it sends frame feed at fixed intervals to the Recognition engine.

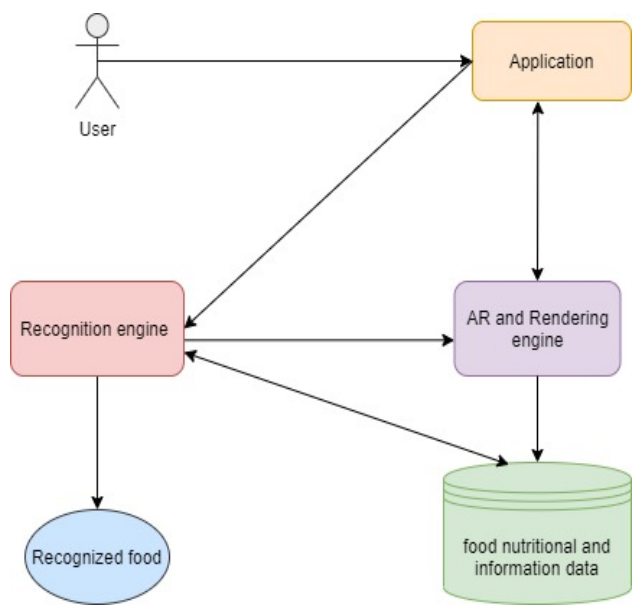

**Fig 4:** Proposed Architecture

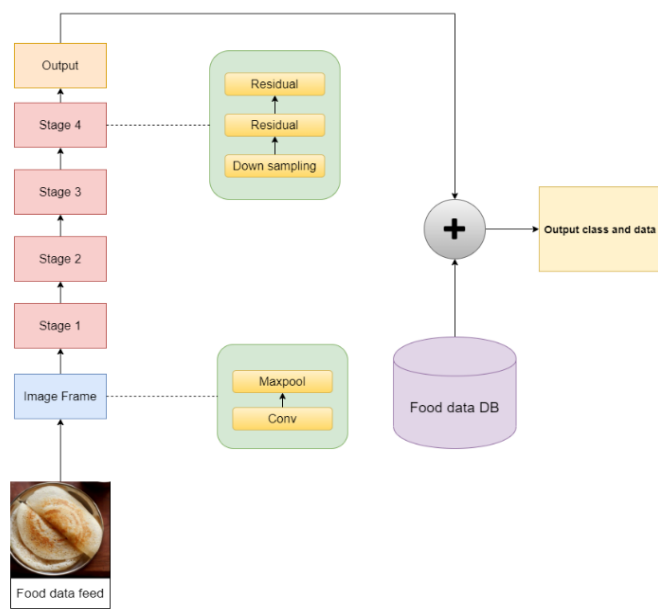

**Fig 5:** Recognition System

The recognition engine classifies the food image into the respective output class, which is then used to find filter the food data, as seen in Fig 5. The resultant class then establishes a connection to the rendering engine, which retrieves nutritional information for the respective class and overlays the data back to the application in realtime.

## 7. Results and Discussions

We have evaluated the classification performance of different ResNet variations on food20 dataset, and this study was differentiated with other renowned architectures, which classified the food into different categories. Figure 6 shows the comparison of the error rate generated by each model over 12 cycles of training, and Figure 7 shows the comparison of Training Loss generated over 12 cycles.

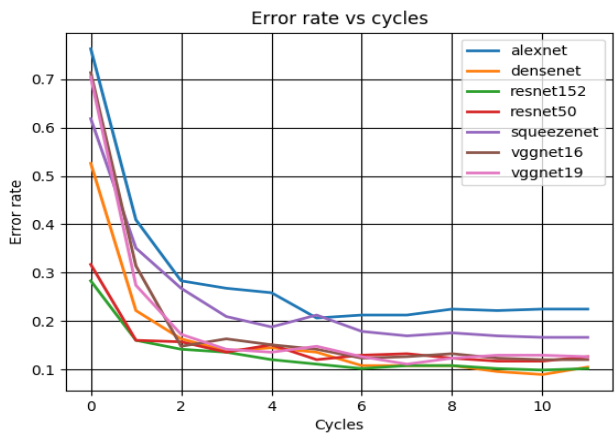 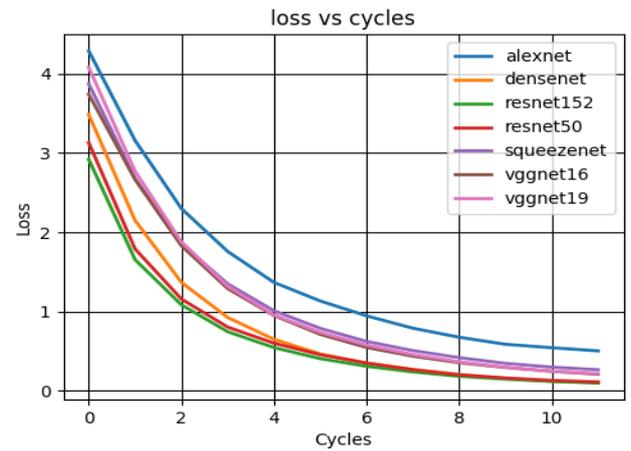

**Fig 6:** Error vs. Cycles for different Models        **Fig 7:** Loss vs. Cycles for different Models

As the inclusion of additional cycles, progress the error rate and losses decrease consequently, Resnet variations clearly outperform the rest from the beginning cycles, It produces the lowest mean error-rate and deviation over 12 classification cycles, as shown in Table 1. It is also to be noted that after six classification cycles, all the models reach a saturation point with minimal deviation in test accuracy. Although Densenet learns to provide a similar accuracy, DenseNets are extremely memory hungry. This is because backpropagation will lead to the accumulation of every layer's outputs, which leads to memory bottleneck as it requires a significant amount of layers and respective output size to function. The ResNet50 that achieves (>90%) is only 50 layers deep, against the 121 layer version of DenseNets used in this experiment, which produces identical results. The main innovation of ResNet is the skip connection. As you know, without adjustments, deep networks often suffer from vanishing gradients, i.e., as the model backpropagates, the gradient gets smaller and smaller. This allows you to stack additional layers and build a deeper network, offsetting the vanishing gradient by allowing your network to skip through layers of it feels they are less relevant in training.

.

| Sno | Model | Dataset | Standard error deviation | Lowest error rate | Mean error rate |
|---|---|---|---|---|---|
| 1 | Resnet50 | Food-20 | 0.052703 | 0.116923 | 0.148718 |
| 2 | Resnet152 | Food-20 | 0.049478 | 0.098462 | 0.130769 |
| 3 | Vgg16 | Food-20 | 0.163696 | 0.120000 | 0.197948 |
| 4 | Vgg19 | Food-20 | 0.159544 | 0.110769 | 0.193333 |
| 5 | Squeezenet | Food-20 | 0.125697 | 0.166154 | 0.239230 |
| 6 | Alexnet | Food-20 | 0.151687 | 0.206154 | 0.292307 |
| 7 | Densenet | Food-20 | 0.115322 | 0.089231 | 0.161794 |

**Table 1:** Error Rates for different Models

Training time to achieve accuracy over 85% have been measured for these models; the results show that residual networks tend to reach higher accuracy in lesser cycles, which in return resulted in lesser training time and resources compared to other models, as displayed in Fig 8. Inference Latency of various models has been compared in Fig 9. A single image frame was supplied to every model several times, and then mean of the inference time for each model was calculated. The same methodology was performed for C.P.U and so for G.P.U. The variations among both tests didn't alter inference time for SqueezeNets, ShuffleNet, and ResNet-50, which produced a lower inference time when distinguished with other models.

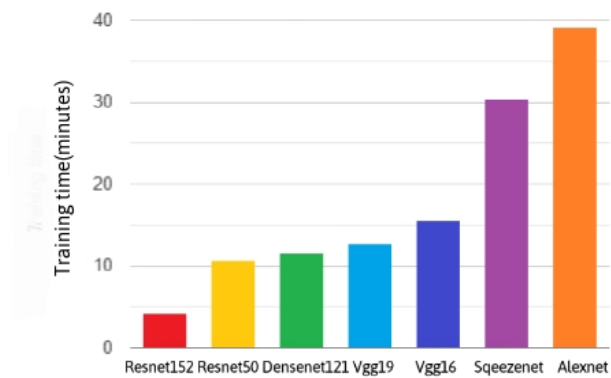

**Fig 8:** Time taken to train the image classifier model to a top-5 validation accuracy of 85% or greater

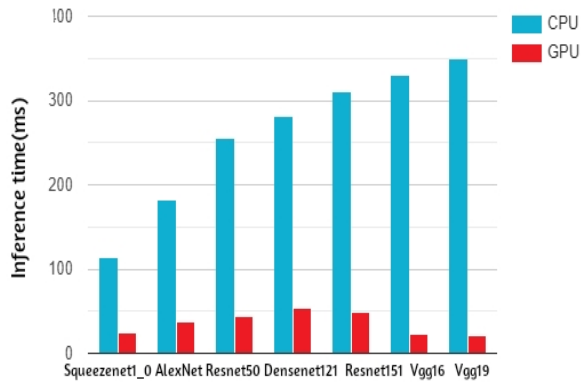

**Fig 9:** Inference Latency: Latency required to classify one image using a model with a top-5 validation accuracy of 93% or greater

We take all three parameters such as Model size, latency, and accuracy and visualize it in a bubble chart, as seen in Figure 10, Bubbles with smaller size are better in aspect of model size, and Bubbles closer to the origin offer a better Speed and Accuracy. Hence with results derived, we can conclude that resnet50 is an optimal model that performs classification optimally and accurately as it covers all three parameters and near to the origin, Densenets performance is comparatively lacking in terms of inference time, squeezenet and alexnet offer poor accuracy.

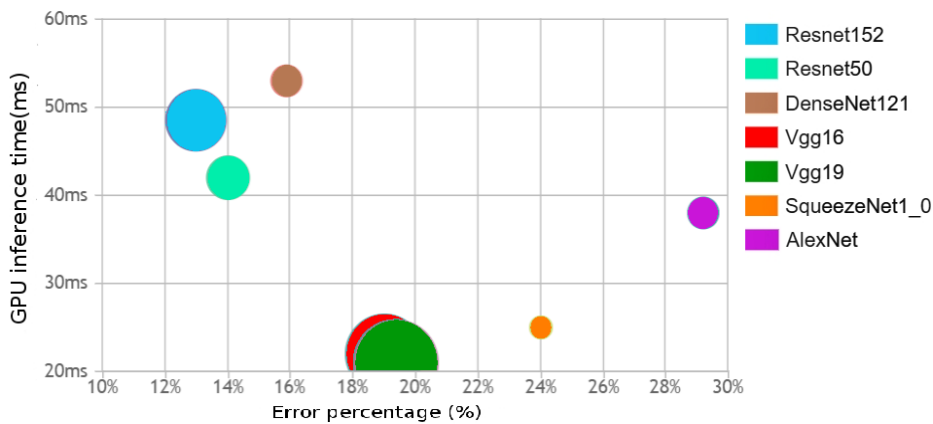

**Fig 10:** Model Comparison based on Accuracy, Speed and Model Size

The classification and information overlay occurred concurrently and rendered as an augmented box and nutritional radar-based graph as follows from the user perspective:

1) The image frames such as Fig 11 are being captured by the application and transferred to retrieve class information:

2) The image class is identified by the recognition system, which then retrieves the class information and passes a flag which is used to trigger a prop in the application, and the food data is then fetched from the Database and rendered into the prop generated concurrently as displayed in Fig 12.

3) Information such as Ingredients, health value generated based on ingredients is retrieved by the AR engine from the Database and invoked when the user decides to view additional information, as displayed in Fig 13.

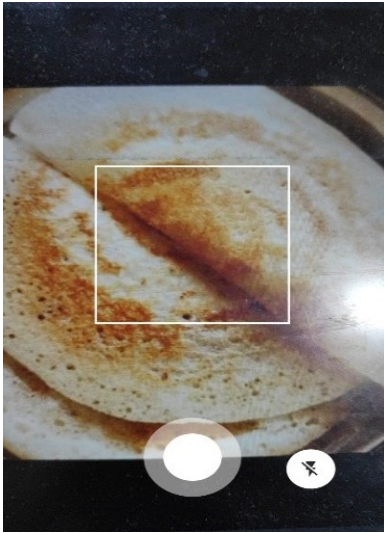

**Fig 11:** Image frame transferred the application

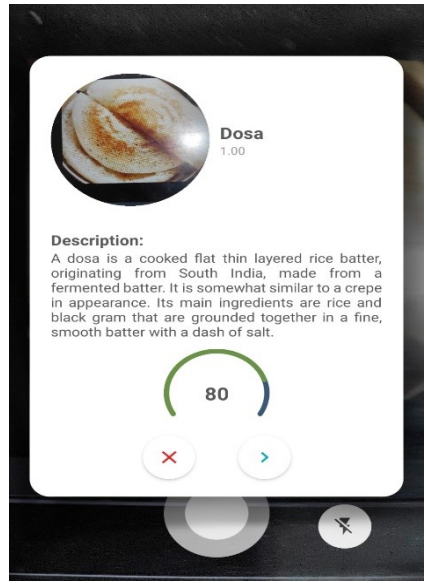

**Fig 12:** Identified Food data retrieved and displayed

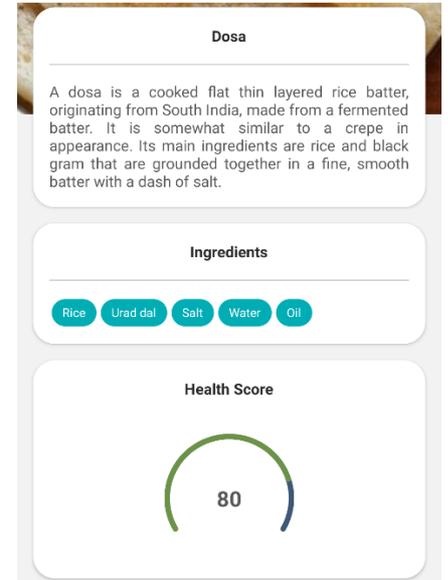

**Fig 13:** Foods information rendered to the user

4) The nutritional data expected from the food class and image default portion set by the user is rendered as a radar chart seen in Fig 14.

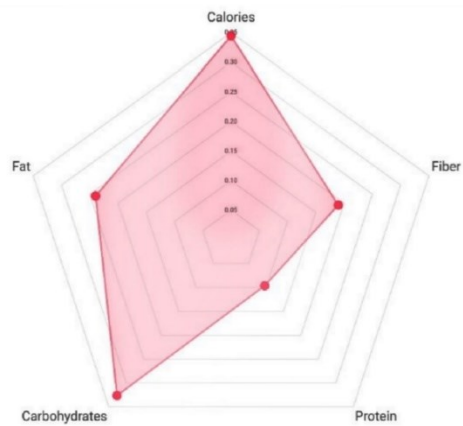

**Fig 14:** Nutritional information rendered to the user

## 8. Conclusions

In this work, We have successfully trained renowned Deep Learning models on our Food-20 dataset. We have observed and visualized the performance of each algorithm and other features for each case and have decided that ResNet is the most suitable model for the application to perform classification both accurately and optimally which we have integrated with an AR-based rendering engine that uses a low complex and instant data rendering procedure. The approach that we used helps retrieve any desired data from the Database optimally and can be used to display the content while the subsequent image frame gets transmitted to the recognition system parallel, which makes it a suitable and more flexible approach for better interactivity and reducing delay. The flexible modular approach we used in the system offers higher quality manageability, scalability, and performance for instant recognition and Rendering.

## 9. Future Work

**Improving the Speed and Accuracy of Residual Networks**

Simplifying ResNet and deducing the backpropagation of the simplified ResNet and incrementing shortcut connections with parameter tuning and optimization strategies to improve the model's speed and accuracy.

**Improved Augmented Reality interactivity**

We have explored and tested food classification on Unity Vuforia AR for much cleaner, and user-friendly interface, Unity framework purpose-built for AR development allowed us to develop the app and deploy it across multiple mobiles and wearable AR devices. Its features include not just core features but unique Unity features that include photorealistic Rendering, physics, device optimizations. Our proposed architecture is yet to be tested on unity to evaluate the performance of cross-compatibility.

## Acknowledgement

I would like to thank Vellore Institute of Technology, Chennai for providing the opportunity and support to work on this paper.

## Notes on Contributors

1. Siddarth Sairaj, is a B.Tech undergraduate student of Computer Science Engineering from Vellore Institute of Technology, Chennai campus. His main fields of interests are Machine Learning, Image Processing, Computer Vision and Augmented Reality.
Email: siddarth.sairaj2017@vitstudent.ac.in
Linkedn: https://www.linkedin.com/in/siddarth-sairaj-41204717a/
Orcid: https://orcid.org/0000-0002-2690-381X

2. Sainath Ganesh, is a B.Tech undergraduate student of Computer Science Engineering from Vellore Institute of Technology, Chennai campus. His main fields of interests are evolutionary algorithms in Artificial Intelligence, Augmented Reality and Virtual Reality in Business.
Email: sainath.g2017@vitstudent.ac.in
Linkedn: https://www.linkedin.com/in/furyswordxd/
Orcid: https://orcid.org/0000-0002-7401-1955

3. Vignesh S, is a B.Tech undergraduate student of Computer Science Engineering from Vellore Institute of Technology, Chennai campus. His main fields of interests are Big Data, Machine Learning, Parallel and distributed computing.
Email: s.vignesh2017a@vitstudent.ac.in
Linkedn: https://www.linkedin.com/in/vignesh-s-035b62179/
Orcid: https://orcid.org/0000-0003-2332-9565

**Footnotes**
Pre-print version of the article is available at https://arxiv.org/abs/2005.04292